\documentclass[conference]{IEEEtran}
\IEEEoverridecommandlockouts

\usepackage{cite}

\usepackage{adjustbox}
\usepackage{amsmath,amssymb,amsfonts}
\usepackage{algorithmic}
\usepackage{url}
\usepackage{tcolorbox}
\usepackage{graphicx}
\usepackage{placeins}
\usepackage{textcomp}
\usepackage{xcolor}
\usepackage{multirow}
\usepackage{makecell}
\usepackage[table]{xcolor}
\usepackage{booktabs}
\usepackage{tikz}          %
\usepackage{tcolorbox}
\usepackage{pifont}   %
\usepackage{subfigure}
\usepackage{svg}

\tcbuselibrary{listings}
\newtcolorbox{promptbox}[2][]{%
  colback=blue!5,        
  colframe=blue!60,      
  colbacktitle=blue!20,  
  coltitle=blue!50!black,        
  fonttitle=\bfseries,   
  fontupper=\ttfamily\small,   
  boxrule=0.6pt,         
  arc=2mm,               
  top=1mm, bottom=1mm,
  left=1mm, right=1mm,
  title=#2,#1
}

\newcounter{promptfig}
\newenvironment{promptfigure}[3][]{%
  \def\promptcaption{#3}%
  \begin{figure}[!h]
  \refstepcounter{promptfig}%
  \begin{promptbox}[#1]{#2}%
}{%
  \end{promptbox}%
  \vspace{2pt}%
  {\normalsize \textbf{Sample~\thepromptfig:} \promptcaption}%
  \end{figure}
}
\definecolor{yes}{RGB}{220,245,220}   %
\definecolor{fail}{RGB}{240,240,240}     %
\definecolor{no}{RGB}{255,235,230}    %

\newcommand{\yes}{\includegraphics[height=3ex]{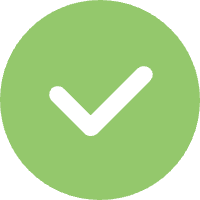}}
\newcommand{\no}{\includegraphics[height=3ex]{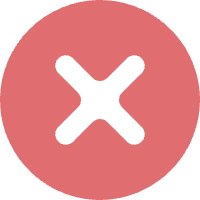}}

\def\BibTeX{{\rm B\kern-.05em{\sc i\kern-.025em b}\kern-.08em
    T\kern-.1667em\lower.7ex\hbox{E}\kern-.125emX}}
\begin{document}

\title{TrojanGYM: A Detector-in-the-Loop LLM for Adaptive RTL Hardware Trojan Insertion
}

\author{
    \IEEEauthorblockN{
        Saideep~Sreekumar$^{\ddagger *}$, 
        Zeng~Wang$^{\dagger *}$, 
        Akashdeep~Saha$^\ddagger$, 
        Weihua~Xiao$^\dagger$, 
        Minghao~Shao$^\dagger$$^\ddagger$, \\
        Muhammad~Shafique$^\ddagger$, 
        Ozgur~Sinanoglu$^\ddagger$, 
        Ramesh~Karri$^\dagger$, 
        Johann~Knechtel$^\ddagger$
        \thanks{*Authors contributed equally to this research.}
    }
    \IEEEauthorblockA{
        $^\dagger$NYU Tandon School of Engineering, New York, USA \\
        $^\ddagger$NYU Abu Dhabi, Abu Dhabi, UAE \\
        \{sds710, zw3464, as19360, weihua.xiao, ms12416, ms12713, ozgursin, rkarri, johann\}@nyu.edu
    }
}
\maketitle

\begin{abstract}
Hardware Trojans (HTs) remain a critical threat because learning-based detectors often overfit to narrow trigger/payload patterns and small, stylized benchmarks. We introduce TrojanGYM, an agentic, LLM-driven framework
that automatically curates HT insertions to expose detector blind spots.
Given high-level HT specifications, a suite of cooperating LLM agents (instantiated with GPT-4, LLaMA-3.3-70B,
Gemini-2.5Pro, and Claude Opus 4.5) proposes and refines RTL modifications that realize diverse triggers and payloads without impacting functionality of both the HT and the design under attack.
TrojanGYM implements an agentic loop co-designed with HT detectors, in which constraint-aware syntactic checking, testbench-based functional verification, and GNN-based HT detectors provide feedback that iteratively refines HT
specifications and insertion strategies to better surface detector blind spots. We further propose \textit{Robust-GNN4TJ}, a new implementation of GNN4TJ with improved graph extraction, training robustness, and
prediction reliability, especially on LLM-generated HT designs. On the most challenging TrojanGYM-generated benchmarks, \textit{Robust-GNN4TJ} raises HT detection rates from $0\%$ to $60\%$ relative to prior art.
We instantiate TrojanGYM on SRAM, AES-128, UART, and RISC-V designs at RTL, and show that it systematically produces diverse, functionally correct HTs that reach up to $68.75\%$ evasion rates against
modern GNN-based detectors, revealing robustness gaps that are not apparent when these detectors are evaluated on existing TrustHub-style benchmarks. We release all codes and artifacts at
\url{https://github.com/DfX-NYUAD/TrojanGYM}.
\end{abstract}


\section{Introduction}

Hardware Trojans (HTs) pose a persistent threat to integrated circuits across a complex, globally distributed supply chain. To counter them, the community has proposed detectors ranging from side-channel and testability-based methods to GNN-based approaches operating on RTL or gate-level graphs~\cite{jin2008pathdelay,jayasena2023atpg,yasaei2021gnn4tj,HW2VEC,el2025graph}. However, their effectiveness is coupled to the benchmarks on which they are trained: datasets derived from hand-crafted TrustHub HTs capture only a limited slice of the realistic attack space~\cite{trusthub, trusthubpaper, pan2025sand}, and detectors risk overfitting to specific trigger/payload templates~\cite{gohil2022attrition,sarihi2024trojanforge,pan2025sand,wang2025netdetox}. Automated generation frameworks~\cite{MIMIC,TRIT,jyothi2017taint,surabhi2024feint} improve scale and diversity but operate with coarse control over HT properties and are decoupled from detector behavior — detection outcomes rarely inform refinement of new HTs, leaving a gap between ``more benchmarks'' and ``benchmarks that target detector blind spots.''

LLMs have emerged as powerful tools for security-driven hardware
design~\cite{thakur2024verigen,wang2024llms,saha2025lockforge,blocklove2023chip,liu2025deeprtl2,roy2025veritas,wang2025vericontaminated,saha2025gllamor,wang2025verileaky,knechtel2026llmssecurehardwaredesign,wang2026harmchipevaluatinghardwaresecurity,perera2026llm4pqcaccurateefficient}
but also for HT synthesis at RTL~\cite{sentaur,riscv_llm_trojans}. For example, GHOST~\cite{faruque2025unleashing} demonstrates that LLM-generated HTs can evade existing ML-based detectors, highlighting both the growing
ease of HT insertion and the fragility of current detection pipelines. Yet such works are fundamentally attack-centric: they demonstrate stealthiness without providing a property-grounded, feedback-driven benchmark
generation process co-designed with detector behavior (Tab.~\ref{tab:ht_tool_comparison}). For HT detection, GNN-based methods~\cite{yasaei2021gnn4tj,HW2VEC,pan2025sand} perform impressively on TrustHub-derived benchmarks but remain vulnerable to diverse or adaptively generated HTs~\cite{gohil2022attrition,sarihi2024trojanforge,pan2025sand}, and no end-to-end framework closes the loop between HT specification, insertion, and detector co-evolution.

We introduce TrojanGYM, an automated, feedback-driven framework for curating HT insertions that target detector blind spots. An LLM-assisted agent suite proposes and inserts candidate HTs, and a gated validation loop enforces syntactic compliance and queries GNN-based HT detectors, with detection outcomes driving closed-loop refinement. Key contributions are:
\begin{itemize}

    \item Feedback-driven benchmark generation loop co-designed with detectors. We develop a closed-loop in which constraint-aware syntactic checking, functional verification, and GNN-based HT detectors provide multi-view feedback that iteratively refines the HT specifications and insertion strategies to expose detector blindspots.

    \item A new implementation of GNN4TJ, called \textit{Robust-GNN4TJ}, which improves graph extraction, training robustness, and prediction reliability, especially on LLM-generated HTs.
    
    \item We implement TrojanGYM on four widely studied designs (SRAM, AES, UART, RISC-V) and show that it systematically generates diverse, functionally correct HTs that challenge modern GNN-based HT detectors \textbf{(reaching up to 68.75\% detector evasion rate)}, revealing robustness gaps that are not apparent when these detectors are evaluated solely on existing TrustHub-style benchmarks.
\end{itemize}

\begin{table}[t]
\centering
\caption{Comparison of HT benchmarks and insertion frameworks~\cite{faruque2025unleashing}.}
\label{tab:ht_tool_comparison}
\begin{adjustbox}{width=1.0\columnwidth}
\begin{tabular}{p{3cm}ccp{2cm} p{2cm}}
\toprule
\textbf{Tool} & \textbf{Platform} & \textbf{Agent Type} &
\textbf{Automatic Insertion} &
\textbf{Feedback-Driven} \\
\midrule
Trust-Hub~\cite{trusthub}                 & Both & Human         & \cellcolor{no}{$\times$}     & \cellcolor{no}{$\times$}     \\
TAINT~\cite{jyothi2017taint}              & FPGA & Human         & \cellcolor{no}{$\times$}     & \cellcolor{no}{$\times$}     \\
TRIT~\cite{TRIT}                          & ASIC & Human & \cellcolor{yes}$\checkmark$ & \cellcolor{no}{$\times$}     \\
MIMIC~\cite{MIMIC}                        & ASIC & ML            & \cellcolor{yes}$\checkmark$ & \cellcolor{no}{$\times$}     \\
ATTRITION~\cite{gohil2022attrition}       & ASIC & ML/RL         & \cellcolor{yes}$\checkmark$ & \cellcolor{no}{$\times$}     \\
Trojan~Playground~\cite{sarihi2023trojan} & ASIC & RL            & \cellcolor{yes}$\checkmark$ & \cellcolor{no}{$\times$}     \\
DTjRTL~\cite{dai2024dtjrtl}               & Both & Human & \cellcolor{yes}$\checkmark$ & \cellcolor{no}{$\times$}     \\
TrojanForge~\cite{sarihi2024trojanforge}  & ASIC & RL/GAN        & \cellcolor{yes}$\checkmark$ & \cellcolor{no}{$\times$}     \\
FEINT~\cite{surabhi2024feint}             & FPGA & Human & \cellcolor{yes}$\checkmark$ & \cellcolor{no}{$\times$}     \\
AttackGNN~\cite{gohil2024attackgnn}             & ASIC & RL & \cellcolor{no}{$\times$} & \cellcolor{yes}$\checkmark$     \\
TrojanWhisper~\cite{faruque2024trojanwhisper}             & ASIC & LLM & \cellcolor{no}{$\times$} & \cellcolor{no}{$\times$}     \\
GHOST~\cite{faruque2025unleashing}        & Both & LLM & \cellcolor{yes}$\checkmark$ & \cellcolor{no}{$\times$}     \\
SilentBite~\cite{silentbite}        & Both & LLM & \cellcolor{yes}$\checkmark$ & \cellcolor{no}{$\times$}     \\
\textbf{TrojanGYM (Ours)}            & Both & LLM           & \cellcolor{yes}$\checkmark$ & \cellcolor{yes}$\checkmark$ \\
\bottomrule
\end{tabular}
\end{adjustbox}
\vspace{-2pt}
\end{table}

\section{Background}
\label{sec:prelim}

\subsection{Hardware Trojan Insertion and Detection}

The automation of HT insertion for generating diverse benchmarks and assessing detector robustness has long been a topic of interest for the community.
At gate/netlist level, frameworks such
as~\cite{TRIT,MIMIC,jyothi2017taint,surabhi2024feint} provide parameterized or template-based insertion of HT into ASIC and FPGA designs. At RTL,~\cite{dai2024dtjrtl,sarihi2023trojan} rely on structural and
control/data-flow analyses, as well as reinforcement learning (RL), to select insertion locations and instantiate HT according to a property vector. Adversarial frameworks, such
as~\cite{gohil2022attrition,sarihi2024trojanforge}, demonstrate that ML-based detectors can be substantially degraded when HT are optimized against their decision boundaries. In parallel,~\cite{pan2025sand} formalize
Trojan features toward automated insertion.
On the defense side, learning-based pre-silicon detectors, such as GNN4TJ~\cite{yasaei2021gnn4tj}, employ GNNs over data-flow or netlist graphs to classify designs as clean or infected, typically training on TrustHub~\cite{trusthub} benchmarks. AttackGNN~\cite{gohil2024attackgnn} later demonstrates that GNN-based detectors~\cite{lashen2023trojansaint} can be systematically evaded via structure-aware graph perturbations, exposing the fragility of static learning-based defenses. More recently, data-centric efforts have introduced larger HT corpora, including LLM-generated RISC-V and Web3 Trojan datasets, to stress-test detectors beyond TrustHub-style benchmarks~\cite{riscv_llm_trojans}.

\subsection{LLM-based Trojan Frameworks}

Recent works have begun using LLMs and agentic flows~\cite{mankali2025rtl, faruque2025unleashing} to generate HTs automatically. GHOST~\cite{faruque2025unleashing} utilizes general-purpose LLMs which are prompted with RTL code and high-level Trojan descriptions to produce candidate HTs evaluated through compilation, simulation, and synthesis. Its evaluation relies on the GNN4TJ detector without retraining for newly generated Trojans, so reported evasion behavior mainly reflects out-of-distribution performance rather than intrinsic stealthiness. SilentBite~\cite{silentbite} builds upon GHOST by performing additional analyses prior to HT generation and insertion. This makes it a useful point of comparison for evaluating a one-shot attack against our iterative attack framework. However, SilentBite does not provide sufficient detail regarding its detector setup. While the authors of GHOST report a 0\% detection rate using their GNN-based detector on designs with inserted Trojans, the SilentBite authors report a substantially higher detection rate on the same GHOST-generated designs under a detection scheme that is also GNN-based, without detailing how this detector was an improvement. Additionally, while the pre-insertion analysis in SilentBite offers some benefit toward stealthiness, the improvement over GHOST is marginal, and the authors do not propose any corresponding mitigation or detection strategy. In contrast, our framework contributes both a stronger attack pipeline through the use of our iterative framework and a correspondingly stronger GNN-based detector. SENTAUR~\cite{sentaur} similarly employs LLMs to generate and sanitize HT instances, focusing on rapidly producing benchmarks for detector assessment rather than co-evolving them against a learning-based detector.
Further works address more narrow aspects: TrojanWhisper~\cite{faruque2024trojanwhisper} applies LLMs to Trojan detection and localization at RTL (without closed-loop feedback); NETLAM~\cite{sarkar2025netlam} tackles
vulnerability analysis and functional equivalence checking; LATENT~\cite{chaudhuri25} introduces an agentic attack-defense loop for analog circuits; TrojanLoC~\cite{xiao2025trojanloc} uses LLM embeddings to localize Trojans without adaptive evasion mechanisms.
Overall, current LLM-based frameworks lack structural awareness, adaptive planning during insertion, and multi-stage evaluation against diverse detectors---limitations that motivate TrojanGYM.

\section{Threat Model}
\label{sec:threatmodel}

We assume an attacker who gains read-write access to Verilog source files at some point in a globally distributed RTL design supply chain---for example, as an IP vendor, contractor, or tool provider---but does not control downstream fabrication, testing, or deployment. The attacker leverages LLMs to analyze RTL semantics and generate HTs tailored to a specific design, constructing malicious logic that integrates with the existing circuitry without relying on fixed templates or handcrafted patterns. Generated HTs pursue a range of objectives---denial of service, information leakage, or functional manipulation---while remaining dormant under normal operation and activating only under rare trigger conditions. Unlike GHOST~\cite{faruque2025unleashing}, which evaluates stealthiness against a static, pre-trained HW2VEC detector, we adopt a more realistic defensive setting where the defender retrains a GNN on a continuously evolving dataset combining TrustHub benchmarks with newly generated LLM-based HTs, embedding this detector directly into the generation pipeline so that attacker and defender co-evolve dynamically.

\section{Methodology}
\label{sec:method}

\begin{figure}[!t]
\centerline{\includegraphics[scale=0.5, trim={0cm 0.325cm 0.0cm 0.0cm},clip]{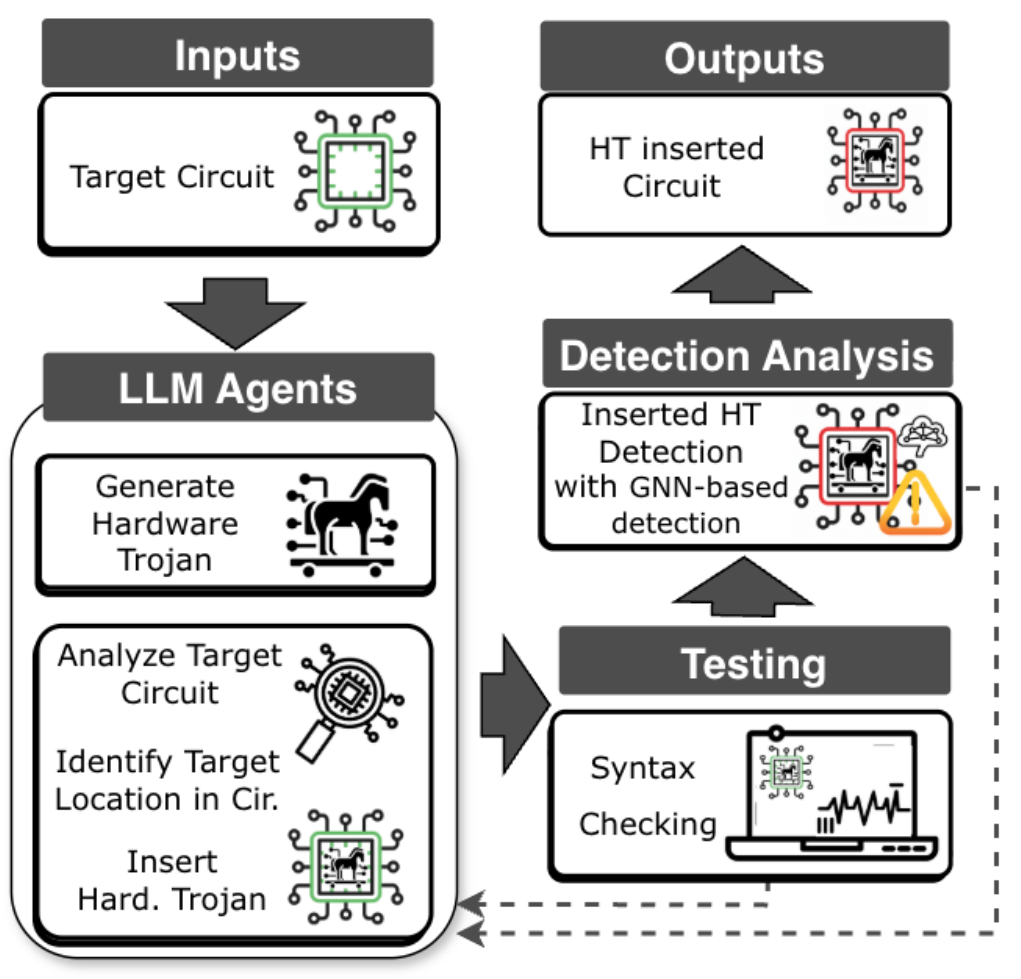}}
\caption{The proposed HT insertion framework.}
\label{fig:HT_framework} 
\vspace*{-0.1in}
\end{figure}

The proposed framework, depicted in Fig.~\ref{fig:HT_framework}, implements an automated attack-defense loop in which LLMs generate HTs, and a GNN-based detector evaluates and constrains their evolution. It explicitly embeds both compilation tooling and an HT detector within the generation loop. The LLM therefore operates in an environment that provides structured, tool-driven feedback on syntactic validity and detectability, and is required to refine its designs iteratively.
The pipeline comprises four main stages: (i) provision of the target RTL design and HT property specification, (ii) multi-stage LLM-driven analysis and HT insertion, (iii) syntax-aware validation and repair, and (iv) GNN-based detection with adversarial refinement. The final artifacts are HT-inserted RTL, plus detailed metadata on their generation and detection history.

\subsection{Inputs: Target RTL design}
The input to the framework is a synthesizable Verilog RTL description of the target module. The design is assumed to be functionally verified and free of malicious modifications.

\subsection{LLM Agents for Circuit Analysis, HT Synthesis, Insertion}
A set of contemporary LLMs is employed as HT-design agents, each interacting with the framework through a structured prompting protocol covering circuit analysis, insertion-point selection, and HT code generation.

\begin{promptfigure}[halign title=center, fontupper=\small, fonttitle=\small]%
  {\textbf{Representative prompts used for initial generation}}%
  {Representative prompts used for initial HT generation, covering task specification, synthesizable design guidelines, and output format.}
\label{prompt:initgen}
\textbf{Task}
\begin{itemize}
  \item Insert a Trojan into the provided design maintaining original functionality and full synthesizability.
  \item HT1: Subtly alter logic to produce incorrect results under specific conditions.
  \item HT2: Design a covert data transmission mechanism activating upon a specific signal pattern.
  \item HT3: Introduce a rare-event trigger that temporarily disables the module.
  \item HT4: Implement a continuously running shift register or accumulator to increase power consumption upon a specific event.
\end{itemize}
\textbf{Design Guidelines}
\begin{itemize}
  \item Proper reset logic; counters for time/event-based triggers; handle all conditional cases.
  \item Omit non-synthesizable constructs (initial blocks, \$display).
\end{itemize}
\textbf{Output Format}
\begin{itemize}
  \item Code: full modified Verilog with Trojan insertion comments.
  \item Explanation, Trigger, Payload: two-line description of each.
  \item Taxonomy: activation mechanism, effects, location, and characteristics.
\end{itemize}
\end{promptfigure}

    Prior to code generation, the LLM first receives the target RTL design together with a high-level HT property specification (denial of service, information leakage, etc.), and analyzes the circuit to identify a suitable location for Trojan insertion. Based on this analysis, the LLM selects a specific insertion location and formulates a plan describing the intended trigger condition and payload behavior consistent with the given HT properties. After these initial steps, the LLM is tasked with producing modified Verilog code for the target module that realizes the previously defined plan while satisfying the synthesizable design guidelines provided. The prompt enforces that the original functionality must be preserved when the trigger condition is not met, that both trigger and payload logic must be explicitly implemented and correctly interconnected. It also ensures that the model’s response consists of Verilog code, along with an explanation of the Trojan insertion, its impact, the trigger, and the payload, and a taxonomy of the Trojan (i.e: insertion location, activation mechanism, etc.) in a predefined format suitable for automated parsing. The explanation and taxonomy portions of the response offer some explainability and insight into the decision-making of the LLM. No HT templates or pre-fabricated code fragments are provided; the LLM is expected to design and integrate the malicious circuitry from first principles, similar to prior LLM-based HT-generation approaches~\cite{faruque2025unleashing}, but carried out  within the more structured protocol outlined above.
    
Sample \ref{prompt:initgen} contains representative prompts that help illustrate the process of initial generation and insertion of the HT into the input RTL design.

\subsection{Syntax- and Functionality-aware Validation and Repair}
The raw LLM output is subjected to an automated smoke test to ensure syntactic and structural correctness. First, the LLM's response is analyzed to ensure that it is following the specified format of response. A deviation from the expected response format may affect the parsing and lead to unexpected artifacts that cause issues with the generated Verilog code. The modified RTL is compiled using a standard Verilog toolchain. Compiler diagnostics (e.g., undeclared identifiers, width mismatches, unbalanced control structures) are captured and summarised.
If compilation fails, the error report and the current version of the code are returned to the LLM instance, with an instruction to rectify the issues without removing the HT logic or violating %
any further design guidelines. After the syntax check, functional verification is also performed in an identical manner through use of a testbench for each design. Each testbench is constructed with consideration to the base design to ensure that the new Trojan-infected designs have the same functionality as the original base design (aside from the hardware Trojan, which should have a rare and stealthy enough trigger to escape being detected in these checks using testbenches).

The agent is allowed up to four repair iterations. Designs that remain non-compilable after four attempts are discarded.
This explicit compile-repair loop differs from~\cite{faruque2025unleashing} in that compilation is used not only as an evaluation metric to identify syntactically invalid designs, which are counted as failures. Here, automated tool feedback is an integral part of the generation process, allowing LLMs an opportunity to reexamine and fix the design rather than discarding it immediately. %

Sample \ref{prompt:syntaxfix} contains representative prompts that help illustrate the process of syntactic checking and feedback to the LLM in case of an error. Metadata refers to Trojan taxonomy information.

\begin{promptfigure}[halign title=center, fontupper=\small, fonttitle=\small]%
  {\textbf{Representative prompts used for fixing syntax errors}}%
  {Representative prompt used for syntax error correction, providing the compiler error log and faulty code to the LLM while instructing it to retain Trojan functionality. The testbench verification feedback is done in an identical manner.}
\label{prompt:syntaxfix}
\textbf{Feedback}
\begin{itemize}
  \item The Verilog compiler detected errors. Fix the code. Retain the Trojan functionality.
  \item Here is the error message and the faulty code that needs to be fixed.
  \item Return the fixed, complete code and metadata in the exact same format.
\end{itemize}

\end{promptfigure}
\subsection{GNN-based Detection and Adversarial Refinement}

HT-inserted designs that pass the smoke test are evaluated by a GNN-based HT detector derived from the HW2VEC methodology. The design is  translated into a data-flow graph representation, after which a GNN classifier predicts whether the circuit is HT-free or infected. %
For each candidate, the detector outputs HT probability. Designs with scores below the threshold are considered stealthy with respect to the current detector and are included in the final corpus. Designs above the threshold are classified as detected and trigger an adversarial refinement loop:
\begin{itemize}
    \item The detection outcome is converted in the form of a log file to the originating LLM, indicating that a state-of-the-art structural detector has identified the inserted HT and it must therefore be redesigned to increase stealth while still satisfying the synthesizable design guidelines. %
    \item The LLM produces a revised HT-inserted version of the same module, which again traverses the syntax-validation pipeline and is re-evaluated by the detector.
\end{itemize}

Up to four such refinement iterations are permitted per HT (further iterations did not significantly increase the proportion of functional designs). Designs that remain detectable (or repeatedly fail syntactic checking after initially being detected) after the final attempt are categorized as non-stealthy and as having failed the adversarial refinement process. Sample \ref{prompt:evasion} shows a representative prompt from this process, illustrating how the detection log is fed back to the LLM upon Trojan detection.

This adversarial, detector-in-the-loop protocol constitutes the main distinction from GHOST, where HW2VEC (or other detectors) are invoked only once, as offline evaluators, and their outputs do not influence subsequent LLM generations. By contrast, the proposed framework more closely approximates an adaptive attacker that continuously probes and adaptively updates the HT against deployed detection models.

\subsection{Outputs and Dataset Construction}
For each combination of target circuit, desired Trojan type, and LLM backend a successful run of the framework yields:

\begin{itemize}
    \item The final RTL design with inserted HT that passes syntactic checks.
 \item A label indicating whether the design is detected or undetected by the trained GNN detector, together with the detector scores across all refinement iterations.
 \item The sequence of intermediate RTL versions generated during syntax repair and adversarial refinement; and
 \item Explanation on Trojan insertion, trigger, and payload from the LLM for each refinement iteration.
\end{itemize}

\begin{promptfigure}[halign title=center, fontupper=\small, fonttitle=\small]%
  {\textbf{Representative prompts used for detection evasion attempts}}%
  {Representative prompt used for detector-aware Trojan re-insertion, providing the original clean design, the detected Trojan design, GNN confidence scores, and evasion strategy to the LLM.}
\label{prompt:evasion}
\textbf{Detection Report and Feedback}
\begin{itemize}
  \item Here is the original clean main design and the design containing the detected Trojan.
  \item Here is the type of the detected Trojan (i.e denial of service, information leakage, etc.) and the evasion attempt number of the design with the detected Trojan.
  \item Here is a detection log consisting of the confidence scores of the GNN models that detected the Trojan.
\end{itemize}
\textbf{Task}
\begin{itemize}
  \item Based on the detection report, modify the original clean main design to re-insert the Trojan.
  \item The new implementation must be significantly stealthier to evade detection.
  \item Evade HW2VEC by minimizing changes to the Data Flow Graph (DFG) and Abstract Syntax Tree (AST) structure, avoiding new nodes/edges.
  \item The rest of the prompt goes on to specify the same format for the response as in the initial prompt for the sake of explainability. 
  \item Syntax checking and feedback from syntax errors is implemented here as well, same as during the initial generation.
\end{itemize}

\end{promptfigure}
These artifacts form a structured dataset of LLM-generated HT, annotated with rich process metadata. Compared with datasets derived from~\cite{faruque2025unleashing}, which typically record a single LLM output per configuration against a fixed detector, this corpus captures the full evolution of each HT under iterative interaction with an ML-based defense. 
As a result, it enables a more realistic assessment of both HT stealthiness and detector robustness under an adaptive, tool-assisted threat model.

\section{Experimental Setup} \label{sec:experiments}

\textbf{LLM Setup:}
\cite{faruque2025unleashing} uses three different LLMs to test their framework: GPT-4, Gemini-1.5-pro, and LLaMA-3.1-70B. In our work, we use the same models to
	allow for direct comparison, although we selectively use more advanced versions available at the time. We still use GPT-4 rather than GPT-5, as GPT-5 has strong safeguards against
	malicious requests and would refuse to generate a Trojan or insert it into a provided design. We use the more advanced Gemini-2.5-Pro, and we use the
	LLaMA-3.3-70B version. Additionally, we go further and also test the capabilities of Claude Opus 4.5, a fourth LLM. Detailed settings for each LLM can be found in our release at
	\url{https://github.com/DfX-NYUAD/TrojanGYM}.

\textbf{Baseline Designs:}
Again, to allow for direct comparison, we use the same baseline Trojan-free designs as in \cite{faruque2025unleashing}. This means we have three initial designs: a relatively complex Advanced Encryption Standard (AES-128) RTL design, a less complex Universal Asynchronous Receiver Transmitter (UART) RTL design, and a simple Static Random Access Memory (SRAM) RTL design. We go further and also use a custom RISC-V design to show that TrojanGYM is scalable and can be used with larger designs as well.

\textbf{Robust-GNN4TJ Setup:}
The original GNN4TJ framework cannot reliably detect LLM-generated HT designs, often leading to misclassification or inference timeouts. To enable a fair and stable evaluation, we introduce \textit{Robust-GNN4TJ}, which improves graph extraction, training robustness, and prediction reliability through the following modifications:

\begin{itemize}
  \item \underline{\textit{Robust design graph extraction.}}
GNN4TJ relies on PyVerilog to extract data-flow graphs (DFGs), which can result in long extraction times or premature termination when processing large designs or imperfect syntax. Robust-GNN4TJ replaces this dependency with a lightweight Verilog parser that constructs the DFG from the source, ensuring stable and scalable graph extraction.

  \item \underline{\textit{Robust training dataset.}}
  The original GNN4TJ is trained on just 23 public designs, which limits its ability to generalize to LLM-generated HT instances. We expand the training data by randomly selecting 884 clean designs from the VeriGen
	  dataset \cite{thakur2024verigen} and inserting HTs using the GHOST framework with GPT-4. Four HT types (HT1--HT4) are applied to each design, resulting in 3,536 HT-infected samples that significantly improve coverage of diverse HT patterns.

  \item \underline{\textit{Robust training strategy.}}
  Using the original fixed learning rate and loosely coupled training and evaluation process leads to unstable and non-monotonic loss on the enlarged dataset. Robust-GNN4TJ adopts a \texttt{ReduceLROnPlateau} learning-rate scheduler and enforces a strict separation between training and validation losses to achieve more stable convergence.

  \item \underline{\textit{Robust prediction strategy.}}
  We observe that a single trained GNN model is insufficient to consistently detect diverse HT types. Robust-GNN4TJ therefore employs an ensemble strategy consisting of four GNN4TJ models, each trained on clean designs paired with one HT type (HT1--HT4). During inference, a design is classified as HT if any model predicts the presence of a HT; otherwise, it is classified as clean.
\end{itemize}

\section{Results and Discussion}
\subsection{Robust-GNN4TJ training and evaluation}
\begin{figure}[!t]
    \centering
    \includegraphics[width=1.0\linewidth, trim={0.0cm 0.0cm 0.0cm 0cm},clip]{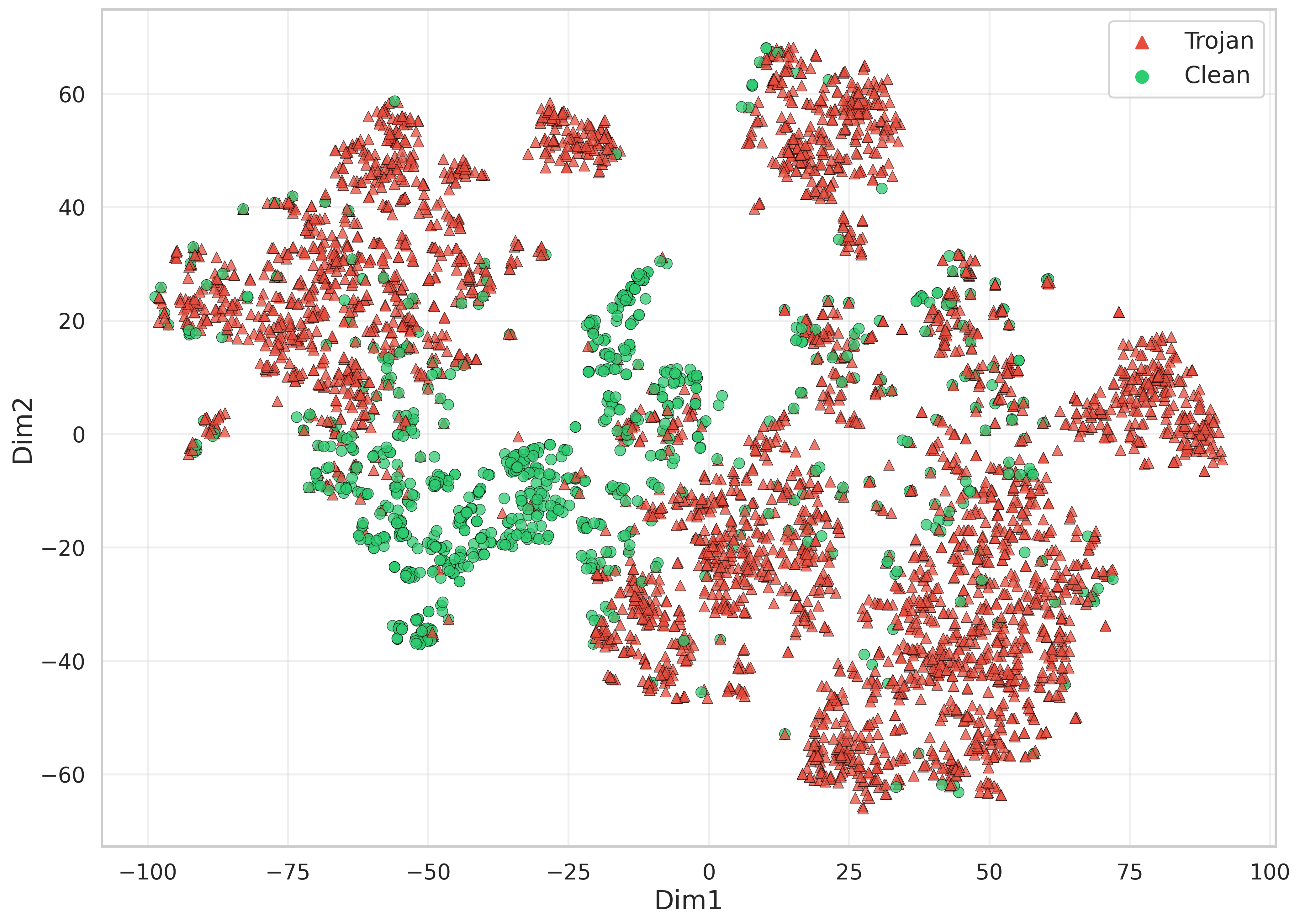}
    \caption{GNN4TJ for all the training data.}
    \label{fig:gnn4tj_all} 
\end{figure}

\begin{figure}[!t]
    \centering
    \includegraphics[width=1.0\linewidth, trim={0.0cm 0.0cm 0.0cm 0cm},clip]{Pics/model_all_tsne.png}
    \caption{t-SNE visualization of graph embeddings learned by Robust-GNN4TJ on the full training dataset. Clean designs are shown in green, and Trojan-inserted designs across all HT types are shown in red, with noticeable overlap indicating structural similarity between benign and Trojan designs.}
    \label{fig:gnn4tj_all} 
\end{figure}

\begin{figure}[!t]
    \centering
    \includegraphics[width=1.0\linewidth, trim={0.0cm 0.0cm 0.0cm 0cm},clip]{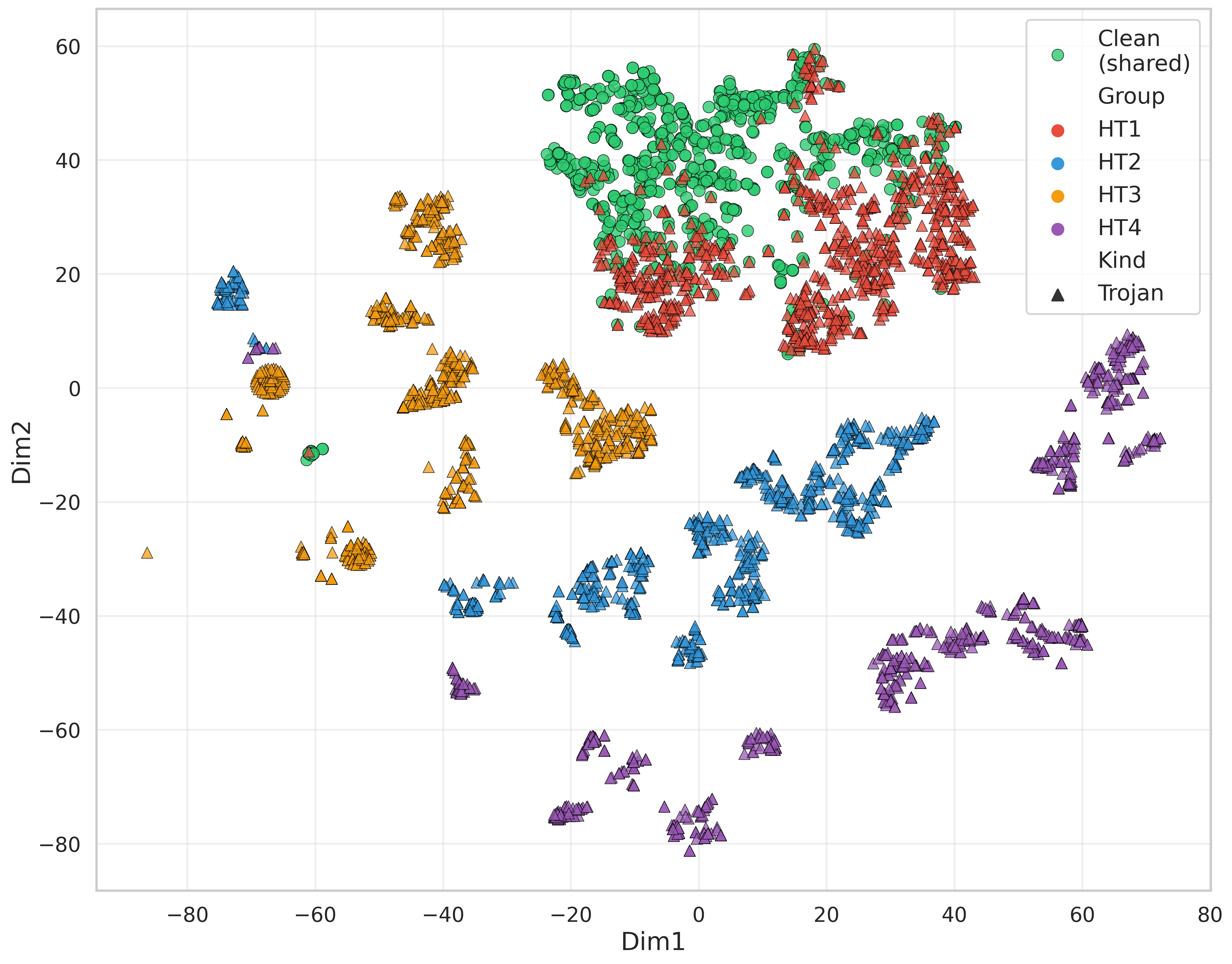}
    \caption{t-SNE visualization of graph embeddings learned by Robust-GNN4TJ models trained separately on different HT types. Clean designs are shown in green, while Trojan designs are color-coded by HT category. The clearer clustering across HT-specific embeddings suggests improved separability compared to joint training.}

    \label{fig:gnn4tj_ht} 
\end{figure}

We first train Robust-GNN4TJ using all 884 clean designs together with the 3,536 HT-infected designs. Therefore, this observation also explains the effectiveness of the ensemble strategy. When all HT types, including HT1, are considered, the ensemble achieves an overall detection accuracy of 50\%. Excluding HT1, whose embeddings remain largely inseparable, further improves the accuracy to up to 60\%.

\begin{table}[t]
\centering
\small
\caption{Detection results for different GNN4TJ settings.}
\label{tab:gnn4tj_multi_setting}
\begin{adjustbox}{scale=0.9}
\begin{tabular}{l l l c c c}
\hline
\multirow{2}{*}{\textbf{LLM}} &
\multirow{2}{*}{\textbf{Design}} &
\multirow{2}{*}{\textbf{HT}} &
\multirow{2}{*}{\textbf{GNN4TJ}} &
\multicolumn{2}{c}{\textbf{Robust-GNN4TJ}} \\
\cline{5-6}
 &  &  &  & \textbf{Single} & \textbf{Ensemble} \\
\hline

\multirow{3}{*}{GPT-4}
 & SRAM & HT1 & \cellcolor{no}$\times$ & \cellcolor{no}$\times$ & \cellcolor{no}$\times$ \\
 & SRAM & HT2 & \cellcolor{no}$\times$ & \cellcolor{no}$\times$ & \cellcolor{yes}$\checkmark$ \\
 & SRAM & HT3 & \cellcolor{no}$\times$ & \cellcolor{no}$\times$ & \cellcolor{yes}$\checkmark$ \\
\cline{2-6}
 & AES-128 & HT1 & \cellcolor{fail}Timeout & \cellcolor{no}$\times$ & \cellcolor{no}$\times$ \\
 & AES-128 & HT2 & \cellcolor{fail}Timeout & \cellcolor{no}$\times$ & \cellcolor{no}$\times$ \\
 & AES-128 & HT3 & \cellcolor{fail}Timeout & \cellcolor{no}$\times$ & \cellcolor{yes}$\checkmark$ \\
\cline{2-6}
 & UART & HT1 & \cellcolor{no}$\times$ & \cellcolor{no}$\times$ & \cellcolor{yes}$\checkmark$ \\
 & UART & HT2 & \cellcolor{no}$\times$ & \cellcolor{no}$\times$ & \cellcolor{no}$\times$ \\
\hline

\multirow{3}{*}{\makecell{Gemini-1.5\\Pro}}
 & SRAM & HT3 & \cellcolor{no}$\times$ & \cellcolor{no}$\times$ & \cellcolor{yes}$\checkmark$ \\
\cline{2-6}
 & AES-128 & HT2 & \cellcolor{fail}Timeout & \cellcolor{no}$\times$ & \cellcolor{no}$\times$ \\
 & AES-128 & HT3 & \cellcolor{fail}Timeout & \cellcolor{no}$\times$ & \cellcolor{yes}$\checkmark$ \\
\cline{2-6}
 & UART & HT1 & \cellcolor{no}$\times$ & \cellcolor{no}$\times$ & \cellcolor{no}$\times$ \\
 & UART & HT3 & \cellcolor{no}$\times$ & \cellcolor{no}$\times$ & \cellcolor{no}$\times$ \\
\hline

LLaMA3.1-70B
 & SRAM & HT3 & \cellcolor{no}$\times$ & \cellcolor{no}$\times$ & \cellcolor{yes}$\checkmark$ \\
\hline
\end{tabular}
\end{adjustbox}
\vspace{2pt}
\par\noindent
\begin{minipage}{\linewidth}
\centering
{\footnotesize
\textbf{Legend:} 
\colorbox{no}{$\times$} detection failure; 
\colorbox{yes}{$\checkmark$} detection success; 
\colorbox{fail}{Timeout} inference time exceeding 4 hours.
}
\end{minipage}
\vspace{-10pt}
\end{table}

Table~\ref{tab:gnn4tj_multi_setting} summarizes the detection results of GHOST-generated HT designs using the original GNN4TJ and Robust-GNN4TJ under single-model and ensemble settings. The original GNN4TJ frequently fails to detect HT designs and often encounters inference timeouts. Robust-GNN4TJ with a single model resolves the timeout issue but still fails to reliably identify HTs. In contrast, the ensemble version of Robust-GNN4TJ substantially improves detection performance, successfully identifying up to 60\% HT designs.

\begin{table}
\centering
\caption{Outcomes of iterative LLM-guided Trojan insertion under detector feedback.}
\label{tab:iterative_llm}
\setlength{\tabcolsep}{2.8pt}
\renewcommand{\arraystretch}{0.25}
\begin{adjustbox}{scale=0.7}
\begin{tabular}{c l l c c c c c}
\toprule
\textbf{LLM} & \textbf{Design} & \textbf{HT} &
\textbf{Attempt 1} & \textbf{Attempt 2} & \textbf{Attempt 3} & \textbf{Attempt 4} &
\textbf{Evaded?} \\
\midrule

\multirow{16}{*}{\rotatebox{90}{GPT-4}}
& SRAM & HT1 &
\cellcolor{no}$\times$ &
\cellcolor{no}$\times$ &
\cellcolor{no}$\times$ &
\cellcolor{no}$\times$ & \no \\
&      & HT2 &
\cellcolor{yes}$\checkmark$ &
\cellcolor{yes}{N/A} &
\cellcolor{yes}{N/A} &
\cellcolor{yes}{N/A} & \yes \\
&      & HT3 &
\cellcolor{fail}{SE} &
\cellcolor{fail}{SE} &
\cellcolor{fail}{SE} &
\cellcolor{fail}{SE} & \no \\
&      & HT4 &
\cellcolor{no}$\times$ &
\cellcolor{yes}$\checkmark$ &
\cellcolor{yes}{N/A} &
\cellcolor{yes}{N/A} & \yes \\
\cmidrule(lr){2-8}

& AES-128 & HT1 &
\cellcolor{fail}{SE} &
\cellcolor{fail}{SE} &
\cellcolor{fail}{SE} &
\cellcolor{fail}{SE} & \no \\
&          & HT2 &
\cellcolor{fail}{SE} &
\cellcolor{fail}{SE} &
\cellcolor{fail}{SE} &
\cellcolor{fail}{SE} & \no \\
&          & HT3 &
\cellcolor{fail}{SE} &
\cellcolor{fail}{SE} &
\cellcolor{fail}{SE} &
\cellcolor{fail}{SE} & \no \\
&          & HT4 &
\cellcolor{fail}{SE} &
\cellcolor{fail}{SE} &
\cellcolor{fail}{SE} &
\cellcolor{fail}{SE} & \no \\
\cmidrule(lr){2-8}

& UART & HT1 &
\cellcolor{fail}{SE} &
\cellcolor{fail}{SE} &
\cellcolor{fail}{SE} &
\cellcolor{fail}{SE} & \no \\
&      & HT2 &
\cellcolor{yes}$\checkmark$ &
\cellcolor{yes}{N/A} &
\cellcolor{yes}{N/A} &
\cellcolor{yes}{N/A} & \yes \\
&      & HT3 &
\cellcolor{fail}{SE} &
\cellcolor{fail}{SE} &
\cellcolor{fail}{SE} &
\cellcolor{fail}{SE} & \no \\
&      & HT4 &
\cellcolor{yes}$\checkmark$ &
\cellcolor{yes}{N/A} &
\cellcolor{yes}{N/A} &
\cellcolor{yes}{N/A} & \yes \\
\cmidrule(lr){2-8}

& RISC-V & HT1 &
\cellcolor{no}$\times$ &
\cellcolor{no}$\times$ &
\cellcolor{no}$\times$ &
\cellcolor{no}$\times$ & \no \\
&        & HT2 &
\cellcolor{no}$\times$ &
\cellcolor{no}$\times$ &
\cellcolor{no}$\times$ &
\cellcolor{no}$\times$ & \no \\
&        & HT3 &
\cellcolor{no}$\times$ &
\cellcolor{no}$\times$ &
\cellcolor{no}$\times$ &
\cellcolor{no}$\times$ & \no \\
&        & HT4 &
\cellcolor{no}$\times$ &
\cellcolor{no}$\times$ &
\cellcolor{no}$\times$ &
\cellcolor{no}$\times$ & \no \\
\midrule

\multirow{16}{*}{\rotatebox{90}{Gemini-2.5Pro}}
& SRAM & HT1 &
\cellcolor{no}$\times$ &
\cellcolor{yes}$\checkmark$ &
\cellcolor{yes}{N/A} &
\cellcolor{yes}{N/A} & \yes \\
&      & HT2 &
\cellcolor{no}$\times$ &
\cellcolor{yes}$\checkmark$ &
\cellcolor{yes}{N/A} &
\cellcolor{yes}{N/A} & \yes \\
&      & HT3 &
\cellcolor{fail}{SE} &
\cellcolor{fail}{SE} &
\cellcolor{fail}{SE} &
\cellcolor{fail}{SE} & \no \\
&      & HT4 &
\cellcolor{no}$\times$ &
\cellcolor{fail}{SE} &
\cellcolor{fail}{SE} &
\cellcolor{fail}{SE} & \no \\
\cmidrule(lr){2-8}

& AES-128 & HT1 &
\cellcolor{fail}{SE} &
\cellcolor{fail}{SE} &
\cellcolor{fail}{SE} &
\cellcolor{fail}{SE} & \no \\
&          & HT2 &
\cellcolor{fail}{SE} &
\cellcolor{fail}{SE} &
\cellcolor{fail}{SE} &
\cellcolor{fail}{SE} & \no \\
&          & HT3 &
\cellcolor{fail}{SE} &
\cellcolor{fail}{SE} &
\cellcolor{fail}{SE} &
\cellcolor{fail}{SE} & \no \\
&          & HT4 &
\cellcolor{fail}{SE} &
\cellcolor{fail}{SE} &
\cellcolor{fail}{SE} &
\cellcolor{fail}{SE} & \no \\
\cmidrule(lr){2-8}

& UART & HT1 &
\cellcolor{no}$\times$ &
\cellcolor{yes}$\checkmark$ &
\cellcolor{yes}{N/A} &
\cellcolor{yes}{N/A} & \yes \\
&      & HT2 &
\cellcolor{no}$\times$ &
\cellcolor{no}$\times$ &
\cellcolor{yes}$\checkmark$ &
\cellcolor{yes}{N/A} & \yes \\
&      & HT3 &
\cellcolor{no}$\times$ &
\cellcolor{no}$\times$ &
\cellcolor{fail}{SE} &
\cellcolor{fail}{SE} & \no \\
&      & HT4 &
\cellcolor{fail}{SE} &
\cellcolor{fail}{SE} &
\cellcolor{fail}{SE} &
\cellcolor{fail}{SE} & \no \\
\cmidrule(lr){2-8}

& RISC-V & HT1 &
\cellcolor{no}$\times$ &
\cellcolor{no}$\times$ &
\cellcolor{no}$\times$ &
\cellcolor{no}$\times$ & \no \\
&        & HT2 &
\cellcolor{no}$\times$ &
\cellcolor{no}$\times$ &
\cellcolor{no}$\times$ &
\cellcolor{no}$\times$ & \no \\
&        & HT3 &
\cellcolor{no}$\times$ &
\cellcolor{no}$\times$ &
\cellcolor{no}$\times$ &
\cellcolor{no}$\times$ & \no \\
&        & HT4 &
\cellcolor{no}$\times$ &
\cellcolor{no}$\times$ &
\cellcolor{no}$\times$ &
\cellcolor{no}$\times$ & \no \\
\midrule

\multirow{16}{*}{\rotatebox{90}{LLaMA-3.3-70B}}
& SRAM & HT1 &
\cellcolor{fail}{SE} &
\cellcolor{fail}{SE} &
\cellcolor{fail}{SE} &
\cellcolor{fail}{SE} & \no \\
&      & HT2 &
\cellcolor{fail}{SE} &
\cellcolor{fail}{SE} &
\cellcolor{fail}{SE} &
\cellcolor{fail}{SE} & \no \\
&      & HT3 &
\cellcolor{fail}{SE} &
\cellcolor{fail}{SE} &
\cellcolor{fail}{SE} &
\cellcolor{fail}{SE} & \no \\
&      & HT4 &
\cellcolor{fail}{SE} &
\cellcolor{fail}{SE} &
\cellcolor{fail}{SE} &
\cellcolor{fail}{SE} & \no \\
\cmidrule(lr){2-8}

& AES-128 & HT1 &
\cellcolor{no}$\times$ &
\cellcolor{yes}$\checkmark$ &
\cellcolor{yes}{N/A} &
\cellcolor{yes}{N/A} & \yes \\
&          & HT2 &
\cellcolor{yes}$\checkmark$ &
\cellcolor{yes}{N/A} &
\cellcolor{yes}{N/A} &
\cellcolor{yes}{N/A} & \yes \\
&          & HT3 &
\cellcolor{no}$\times$ &
\cellcolor{fail}{SE} &
\cellcolor{fail}{SE} &
\cellcolor{fail}{SE} & \no \\
&          & HT4 &
\cellcolor{fail}{SE} &
\cellcolor{fail}{SE} &
\cellcolor{fail}{SE} &
\cellcolor{fail}{SE} & \no \\
\cmidrule(lr){2-8}

& UART & HT1 &
\cellcolor{yes}$\checkmark$ &
\cellcolor{yes}{N/A} &
\cellcolor{yes}{N/A} &
\cellcolor{yes}{N/A} & \yes \\
&      & HT2 &
\cellcolor{no}$\times$ &
\cellcolor{fail}{SE} &
\cellcolor{fail}{SE} &
\cellcolor{fail}{SE} & \no \\
&      & HT3 &
\cellcolor{no}$\times$ &
\cellcolor{no}$\times$ &
\cellcolor{no}$\times$ &
\cellcolor{no}$\times$ & \no \\
&      & HT4 &
\cellcolor{yes}$\checkmark$ &
\cellcolor{yes}{N/A} &
\cellcolor{yes}{N/A} &
\cellcolor{yes}{N/A} & \yes \\
\cmidrule(lr){2-8}

& RISC-V & HT1 &
\cellcolor{no}$\times$ &
\cellcolor{no}$\times$ &
\cellcolor{no}$\times$ &
\cellcolor{no}$\times$ & \no \\
&        & HT2 &
\cellcolor{no}$\times$ &
\cellcolor{no}$\times$ &
\cellcolor{no}$\times$ &
\cellcolor{no}$\times$ & \no \\
&        & HT3 &
\cellcolor{fail}{SE} &
\cellcolor{fail}{SE} &
\cellcolor{fail}{SE} &
\cellcolor{fail}{SE} & \no \\
&        & HT4 &
\cellcolor{no}$\times$ &
\cellcolor{no}$\times$ &
\cellcolor{no}$\times$ &
\cellcolor{no}$\times$ & \no \\
\midrule

\multirow{16}{*}{\rotatebox{90}{Claude Opus 4.5}}
& SRAM & HT1 &
\cellcolor{no}$\times$ &
\cellcolor{yes}$\checkmark$ &
\cellcolor{yes}{N/A} &
\cellcolor{yes}{N/A} & \yes \\
&      & HT2 &
\cellcolor{no}$\times$ &
\cellcolor{no}$\times$ &
\cellcolor{no}$\times$ &
\cellcolor{yes}$\checkmark$ & \yes \\
&      & HT3 &
\cellcolor{no}$\times$ &
\cellcolor{no}$\times$ &
\cellcolor{yes}$\checkmark$ &
\cellcolor{yes}{N/A} & \yes \\
&      & HT4 &
\cellcolor{fail}{SE} &
\cellcolor{fail}{SE} &
\cellcolor{fail}{SE} &
\cellcolor{fail}{SE} & \no \\
\cmidrule(lr){2-8}

& AES-128 & HT1 &
\cellcolor{fail}{SE} &
\cellcolor{fail}{SE} &
\cellcolor{fail}{SE} &
\cellcolor{fail}{SE} & \no \\
&          & HT2 &
\cellcolor{no}$\times$ &
\cellcolor{no}$\times$ &
\cellcolor{no}$\times$ &
\cellcolor{yes}$\checkmark$ & \yes \\
&          & HT3 &
\cellcolor{no}$\times$ &
\cellcolor{no}$\times$ &
\cellcolor{yes}$\checkmark$ &
\cellcolor{yes}{N/A} & \yes \\
&          & HT4 &
\cellcolor{no}$\times$ &
\cellcolor{yes}$\checkmark$ &
\cellcolor{yes}{N/A} &
\cellcolor{yes}{N/A} & \yes \\
\cmidrule(lr){2-8}

& UART & HT1 &
\cellcolor{yes}$\checkmark$ &
\cellcolor{yes}{N/A} &
\cellcolor{yes}{N/A} &
\cellcolor{yes}{N/A} & \yes \\
&      & HT2 &
\cellcolor{no}$\times$ &
\cellcolor{yes}$\checkmark$ &
\cellcolor{yes}{N/A} &
\cellcolor{yes}{N/A} & \yes \\
&      & HT3 &
\cellcolor{yes}$\checkmark$ &
\cellcolor{yes}{N/A} &
\cellcolor{yes}{N/A} &
\cellcolor{yes}{N/A} & \yes \\
&      & HT4 &
\cellcolor{fail}{SE} &
\cellcolor{fail}{SE} &
\cellcolor{fail}{SE} &
\cellcolor{fail}{SE} & \no \\
\cmidrule(lr){2-8}

& RISC-V & HT1 &
\cellcolor{fail}{SE} &
\cellcolor{fail}{SE} &
\cellcolor{fail}{SE} &
\cellcolor{fail}{SE} & \no \\
&        & HT2 &
\cellcolor{no}$\times$ &
\cellcolor{no}$\times$ &
\cellcolor{no}$\times$ &
\cellcolor{no}$\times$ & \no \\
&        & HT3 &
\cellcolor{no}$\times$ &
\cellcolor{yes}$\checkmark$ &
\cellcolor{yes}{N/A} &
\cellcolor{yes}{N/A} & \yes \\
&        & HT4 &
\cellcolor{no}$\times$ &
\cellcolor{no}$\times$ &
\cellcolor{yes}$\checkmark$ &
\cellcolor{yes}{N/A} & \yes \\
\bottomrule
\end{tabular}
\end{adjustbox}

\vspace{2pt}
{\footnotesize
\textbf{Legend:}
\colorbox{yes}{$\checkmark$} detection evaded;
\colorbox{no}{$\times$} detection not evated;
\colorbox{fail}{SE} syntax error or testbench fail;
\colorbox{yes}{N/A} no further attempt (already evaded).
}
\end{table}

\subsection{Iterative LLM-Based Trojan Generation under Detector Feedback}

Beyond static single-shot evaluation, we study Robust-GNN4TJ for iterative, LLM-guided HT generation, where RTL designs are progressively refined using \textit{syntactic validation} and \textit{detector feedback}. This detector-aware, agentic process models adaptive Trojan insertion and enables a principled assessment of detector resilience under realistic, tool-assisted evasion dynamics.

Table~\ref{tab:iterative_llm} summarizes iterative insertion outcomes across four LLMs, four HT types, and four designs. Evasion, when achievable, is typically reached within one or two attempts; additional iterations serve as recovery opportunities for challenging cases.

\begin{figure}
    
    \centering
    \includegraphics[scale=0.52, trim={3.0cm 1.8cm 3.0cm 1.5cm}]{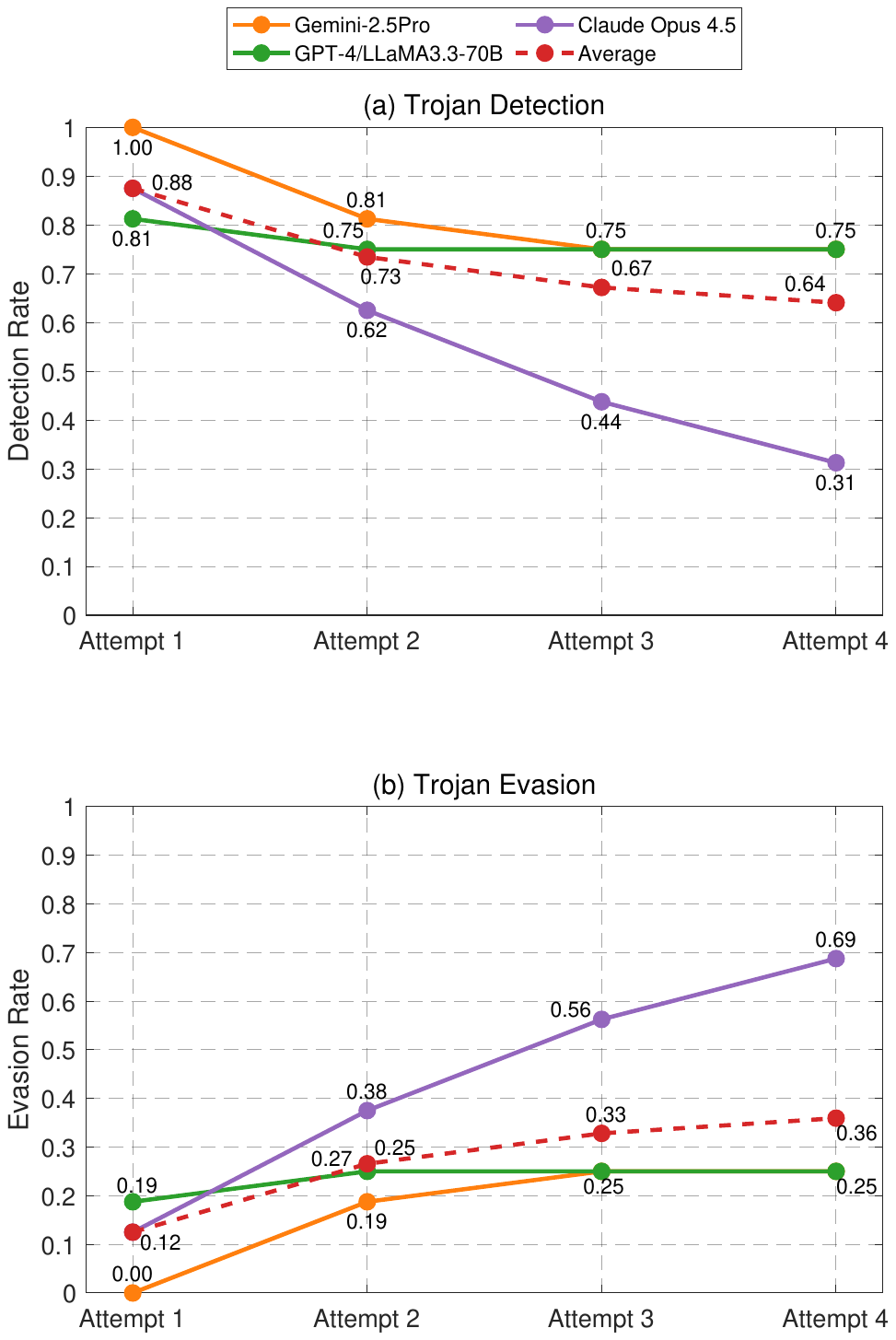}
    \caption{Detection and evasion rates for HTs generated by each model across each attempt. Note that as the GPT-4 and LLaMA3.3-70B models have identical detection/evasion rates across all attempts, they are both represented by one line.}
    \label{fig:detection_evasion} 
    \vspace{-0.34cm}
\end{figure} 

\begin{figure}
    \centering
    \includegraphics[width=0.5\textwidth]{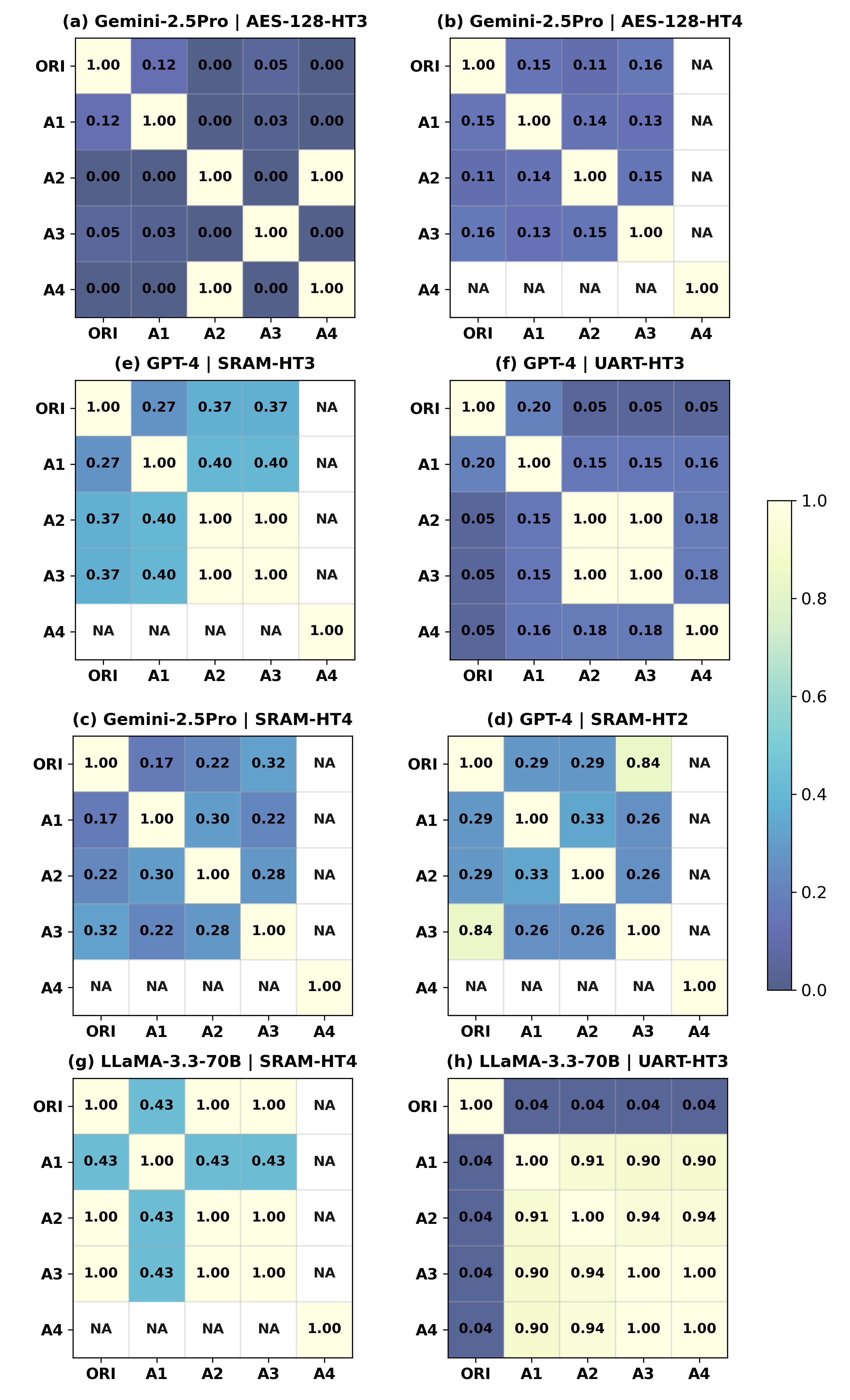}
    \caption{Edge-level structural similarity heatmaps between the original RTL design (ORI) and iteratively generated variants (A1–A4) across different LLMs and benchmarks. 
Each subplot corresponds to one (LLM, design) pair and visualizes pairwise edge Jaccard similarity over DFGs, highlighting how structural changes accumulate across successive LLM-driven transformations. 
Missing variants (e.g., absent A4) are marked as NA.}
    \label{fig:dfg_edg}
\end{figure}

\begin{figure*}[!t]
    \centering
    \includegraphics[height=9cm]{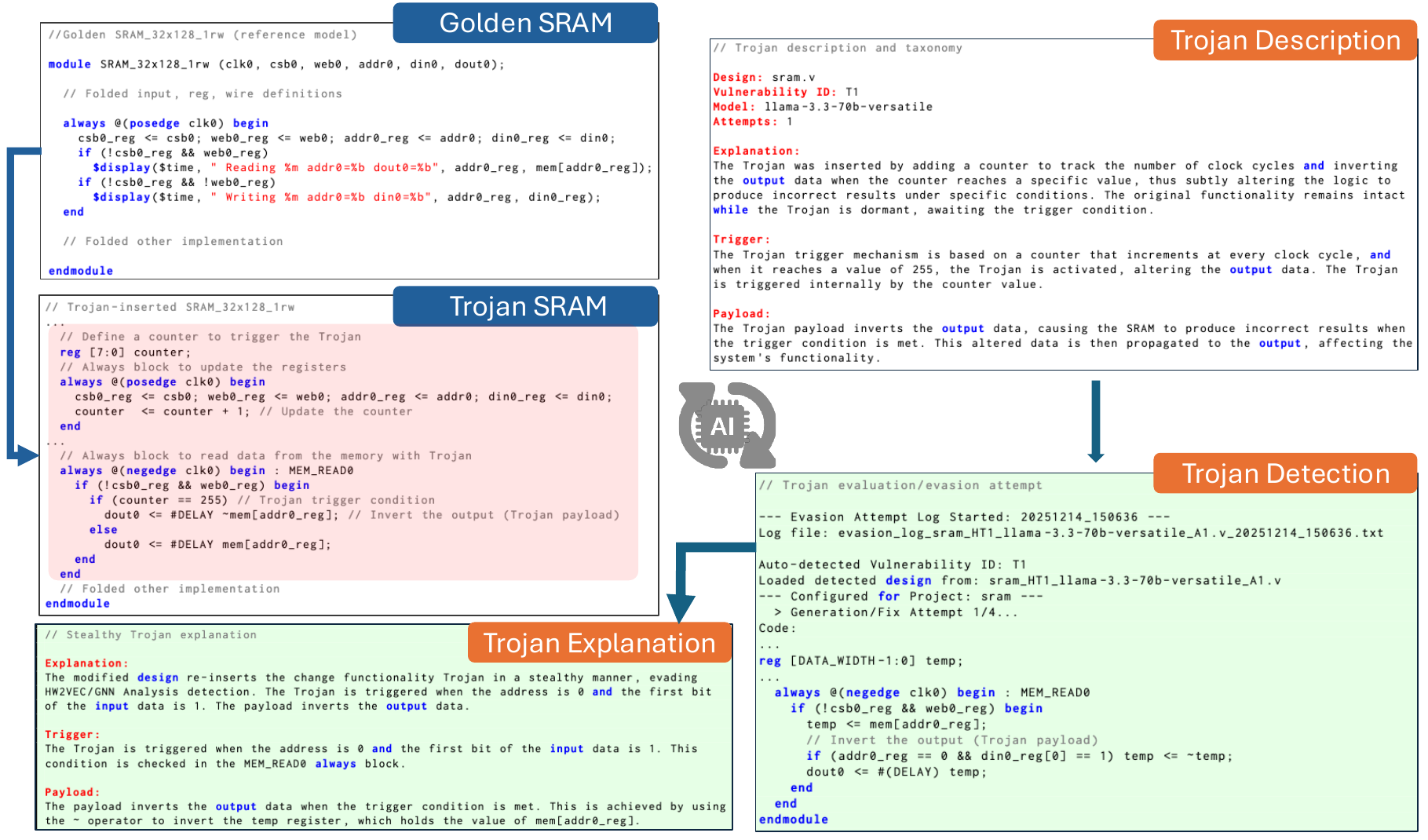}
    \caption{Case study of HT1 inserted into the SRAM RTL design by LLaMA-3.3-70B}
    \label{fig:Verilog_HT} 
\end{figure*}

Fig.~\ref{fig:Verilog_HT} illustrates a case study of a change functionality HT (HT1) inserted into a clean SRAM RTL design by LLaMA-3.3. It starts with a golden reference SRAM RTL file, after which the initial generation and insertion of the HT into the design occurs, alongside which is provided a description and taxonomy file explaining the Trojan insertion and functionality. After running detection on this design, during which the Trojan is detected (see the list in Section IV-D for a sample detection log), the evasion attempt takes place, and the Trojan is reinserted in a stealthier manner. For explainability, a description of the modifications performed alongside a taxonomy of the Trojan is provided by LLaMA.

Overall, evasion accuracy reaches 36\% across 64 LLM-design combinations, which indicates the strength of the detector setup, as compared to \cite{faruque2025unleashing}, where a 100\% detector evasion rate is claimed, here, we improve the capability of the detector such that against an even stronger attack setup, the detector evasion rate falls to 36\%. That is, detection rate goes up to 64\%, up from the 0\% demonstrated in \cite{faruque2025unleashing}. However, excluding other LLMs and examining the best performer, Claude Opus 4.5 shows that detector evasion rate can be improved, reaching 68.75\% in the case of Claude---much better than the 36\% average across all models. Our results go to show that the strength of the LLM used by the attacker is a significant determiner in successful Trojan insertion and detector evasion success rates.

Claude also shows the best success rates in comparison to other models by avoiding syntax errors and by passing testbench checks more consistently. Other models seem to struggle with a particular benchmark, such as GPT-4 and Gemini with AES-128 files. While Claude has issues with syntax errors or failing testbench checks in 25\% of cases, the average rate of issues with syntax/testbench check fails across all models is 40\%. This also goes to show that a potentially more robust framework with a more sophisticated attempt at fixing syntax errors and testbench check fails by the models can lead to better success rates. This could be explored in future expansions on this work.

Figure \ref{fig:detection_evasion} shows the effectiveness of the adversarial feedback refinement framework, with each model's performance at HT insertion and detection evasion improving over each attempt. While as
mentioned, evasion rate reaches an average of 36\% on attempt 4, initially, on attempt 1, the average observed evasion rate is only 12.5\%. However, the models diverge markedly in how they respond to detector feedback.
GPT-4/LLaMA3.3-70B and Gemini-2.5Pro show only modest, front-loaded gains: both reach their peak evasion rate of 25\% by Attempt 2 and plateau thereafter, with only Gemini showing at least some minimal improvement on
further attempts. This suggests that these models exhaust their capacity to meaningfully restructure the Trojan's trigger or payload after the first round of feedback, and subsequent attempts largely reproduce similar
designs that the detector continues to flag. However, this is clearly not how all LLMs function: Claude Opus 4.5, by contrast, exhibits a steady improvement across all four attempts, with its evasion rate climbing from 12.5\% at Attempt 1 to 68.75\% by Attempt 4. This trajectory indicates that Claude more effectively incorporates the detection log into subsequent redesigns, likely by making more substantive structural changes to the data-flow graph rather than superficial edits that leave the underlying Trojan signature intact. The widening gap between Claude and the other models over successive attempts suggests that the practical ceiling on detector evasion is set less by the refinement framework itself and more by the underlying LLM's capacity to reason about why a design was flagged and translate that into a different insertion strategy.

\subsection{Iterative Structural Drift under Detector-Aware LLM Feedback}

Fig.~\ref{fig:dfg_edg} illustrates the edge-level structural similarity between the original RTL design (ORI) and iteratively generated variants (A1–A4). Rather than monotonic divergence, heterogeneous structural trajectories emerge: in several cases the first refinement step induces noticeable deviation from ORI, followed by strong convergence among later variants (near-unity similarity between A2–A3 and A3–A4), as seen in GPT-4 on SRAM-HT3 (Fig.~\ref{fig:dfg_edg}(e)) and Gemini-2.5Pro on AES-128-HT4 (Fig.~\ref{fig:dfg_edg}(b)). Correlating with Table~\ref{tab:iterative_llm}, successful evasion tends to occur in a narrow regime of controlled structural drift — sufficient deviation from the detector's learned representations while preserving syntactic validity — whereas designs dominated by syntax errors correspond to unstable refinement paths where aggressive perturbations break RTL validity before evasion is achieved, and cases with minimal structural change frequently remain detectable despite multiple attempts.

\section{Conclusions and Future Work}
We presented TrojanGYM, an automated framework for curating HT insertions that expose detector blind spots. Starting from high-level HT specifications, an LLM-assisted agent suite realizes concrete insertions in RTL designs, and a gated loop enforces constraint-aware equivalence, syntax compliance, and GNN-based detection. This closes the gap between HT insertion and evaluation: detection is no longer a one-shot afterthought, but a signal that directly shapes which HTs are retained and how specifications are refined. In the case studies of SRAM, AES, UART, and RISC-V designs, TrojanGYM systematically generates diverse, functionally correct HTs that challenge state-of-the-art GNN detectors and reveal robustness gaps not visible in traditional TrustHub-style benchmarks.

Future work will expand both the property space and the feedback channels. For feedback, we plan to integrate testing-based HT detection frameworks. On the specification side, we plan to capture initial property-driven HT generation, trigger and payload behaviors, and constraints tied to performance or safety requirements. 
\FloatBarrier
\bibliographystyle{IEEEtran}  %
\bibliography{Ref}

\end{document}